\documentclass[amsmath,amssymb,aps,prb,showpacs,a4paper,twocolumn,10pt]{revtex4-1}

\usepackage[cp1250]{inputenc} 
\usepackage[T1]{fontenc} 
\usepackage{lmodern}
\usepackage{dsfont, graphics}
\usepackage[usenames,dvipsnames]{color}
\usepackage{braket}

\usepackage{graphicx, epsfig}

\input epsf

\def\beq{\begin{equation}}
\def\eeq{\end{equation}}
\def\beqa{\begin{eqnarray}}
\def\eeqa{\end{eqnarray}}

\newcommand{\eq}[1]{Eq.(#1)}
\newcommand{\fig}[1]{Fig.(#1)}

\newcommand{\col}[1]{\color{black} #1}

\begin{document}

\title{From antiferromagnetic order to magnetic textures in the two dimensional Fermi Hubbard model with synthetic spin orbit interaction}

\author{Ji\v{r}í Miná\v{r}}
\affiliation{Centre for Quantum Technologies, National University of Singapore, Singapore}
\author{Beno\^{i}t Gr\'{e}maud}
\affiliation{Centre for Quantum Technologies, National University of Singapore, Singapore}
\affiliation{Department of Physics, National University of Singapore, Singapore}
\affiliation{Laboratoire Kastler Brossel, Ecole Normale Sup\'{e}rieure CNRS, UPMC; 4 Place Jussieu, 75005 Paris, France}


\begin{abstract}
We study the interacting Fermi-Hubbard model in two spatial dimensions with synthetic gauge coupling of the  
spin orbit Rashba type, at half-filling. Using real space mean field theory, we numerically determine the phase 
as a function of the interaction strength for different values of the gauge field parameters. For a fixed value of the gauge field, we observe 
that when the strength of the repulsive interaction is increased, the system enters into an antiferromagnetic phase, 
then undergoes a first order phase transition to an non collinear magnetic phase. Depending on the gauge field parameter, 
this phase further evolves to the one predicted from the effective Heisenberg model obtained in the limit of large 
interaction strength. We explain the presence of the antiferromagnetic phase at small interaction from the computation 
of the spin-spin susceptibility which displays a divergence at low temperatures for the antiferromagnetic ordering. 
We discuss, how the divergence is related to the nature of the underlying Fermi surfaces. Finally, the fact that the 
first order phase transitions for different gauge field parameters occur at unrelated critical interaction strengths 
arises from a Hofstadter-like situation, i.e. for different magnetic phases, the mean-field Hamiltonians have 
different translational symmetries.	
\end{abstract}

\maketitle

\section{Introduction}

The recent progress of experiments using cold atomic gases\cite{Lewenstein07,Blochreview08,Ketterle2}, in particular in implementing artificial gauge 
fields~\cite{Spielman09a,Spielman09b,Spielman11,Zhang12,Zwierlein12,Windpassinger12,Spielman12}, 
has open the door to the studies of a whole class of model Hamiltonians, some directly inherited from 
condensed matter physics (quantum Hall effects), but, more saliently, some are genuily generating new physical situations, allowing
physicists to further develop and test theoretical ideas, like topological phases~\cite{Mottonen09,Bercioux11}, 
non-abelian particles~\cite{Burrello10} or  mixed dimensional systems~\cite{Nishida_2008,Nishida_2010,Lamporesi_2010,Huang_2013,Iskin_2010}. In particular,
the experiments involving spinors, made of either bosons or fermions in different Zeeman sub-levels,
are now able to produce non-abelian gauge-fields, leading to a kinetic term allowing for a modification of the internal
degrees of freedom along the propagation of the particle~\cite{Dalibard11}. 
In two-dimensional lattices, the non-abelian gauge fields 
result in tight-binding Hubbard models with spin-flip hoping terms, i.e. the hoping matrices are not diagonal anymore in the spin degrees of
freedom. Among all the possibilities, an artificial gauge-field mimicking a spin-orbit coupling term~\cite{Dalibard11} (see below)
is probably the most well-known and studied situation, for two reasons: (i) it corresponds to a spatially independent  vector potential leading to
relatively simple analytical treatment, especially in the bulk situation; (ii) it leads to highly non-trivial features like broken time-reversal
ground states and/or magnetic textures with topological properties, like skyrmion crystal~\cite{Cole_2012,Hofstetter12}. 

The cases of two components bosons or fermions in the presence of a spin-orbit coupling have been the objects of recent analytical and numerical 
studies~\cite{Lewenstein09,Cole_2012,Cai12,Galitski12,Hofstetter12,Fujimoto09,Goldman12}. 
In particular, in the case of repulsive interactions, they have emphasized various magnetic ordering and textures depicted by the effective spins. One
must note that in the large interaction strength, and close to half-filling, both bosonic and fermionic situations can be described by effective 
and quite similar Heisenberg models involving the spin degrees of freedom only (see below). The case of fermions with attractive interaction
has also been studied, 
emphasizing the impact of the spin-flip terms on the pairing states, like the BCS-BEC crossover or the Fulde-Ferrel-Larkin-Ovchinnikov (FFLO) phase
\cite{Shenoy11,Sademelo12,Lewenstein10,Iskin11,Iskin12,Iskin13}. The instabilities of attracting bosons with such a spin-orbit coupling have also been
considered~\cite{Riedl13}.

Even though the Bose-Hubbard model and the Fermi-Hubbard model with a spin-orbit coupling  depicts similar phases 
in the strong (repulsive) interaction limit, the behavior for small coupling is obviously different: the Mott regime of the 
bosonic models turns into a complex superfluid regime whereas fermionic models are expected to be in a Mott insulator state. Still, in the later case,
the evolution of the magnetic ordering from the non-interacting situation towards the Heisenberg model regime remains largely unexplored.
In the present paper, using both a linear response approach and real space mean-field theory, we emphasize
that the Fermi-Hubbard model at half-filling, in the presence of a spin-orbit coupling, still depicts, at low
interaction,  an antiferromagnetic phase, corresponding to the Fermi-Hubbard model \textit{without} spin-orbit coupling. 
The impact of the gauge-field then results, at intermediate interaction depending on the strength of the spin-orbit coupling, 
in a first order phase transition towards a non collinear magnetic order, which 
further evolves to magnetic texture at  large interaction, predicted by the effective Heisenberg model. 

The paper consists in two main parts: in Section~\ref{method}, we describe the theoretical framework (Fermi-Hubbard model, linear response theory, effective
Heisenberg model). In Section~\ref{result} (i) we show that the non-interacting energies (band structure) always 
depict a nesting at the antiferromagnetic order
resulting in a diverging spin-spin susceptibility at low temperature and, thereby, an instability towards an 
antiferromagnetic phase at small interaction; (ii) we provide the numerical mean-field results for different values of the spin-orbit 
coupling and the interaction, showing evidences for the transition from the antiferromagnetic order to a non-collinear magnetic order. In particular,
the skyrmionic phase is shown to already exist for moderate values of the interaction.

\section{Methods}
\label{method}

\subsection{model}

In this paper we consider a system of spin 1/2 fermionic particles in 2D square lattice 
and subjected to a synthetic gauge field (more will be said about the specific form of the gauge field later). 
The lattice spacing is equal to unity, setting both spatial and momentum scales.
The system Hamiltonian reads
\beq
	H_{\rm tot} = H_{\rm kin} + H_{\mu} + H_{\rm int},
\eeq
where
\begin{subequations}
 \begin{align}
	H_{\rm kin} &= - \sum_{\bf j,j'} c^\dag_{{\bf j},s} T^{s, s'}_{\bf j, j'}  c_{{\bf j'},s'} \label{eq H parts kin}  \\
	H_\mu &= - \sum_{\bf j} \mu_1 n_{{\bf j},1} + \mu_2 n_{{\bf j},2} \label{eq H parts mu} \\
	H_{\rm int} &= U \sum_{\bf j} \left(n_{{\bf j},1}-\frac{1}{2}\right)
\left( n_{{\bf j},2}-\frac{1}{2}\right), \label{eq H parts int}	
 \end{align}
\end{subequations}
where $T$ are the tunneling matrices and are taken to be time independent in the following.  $c^\dag_{{\bf j},s}, c_{{\bf j}',s'}$ 
are the usual fermionic creation and annihilation operators satisfying 
$\{ c_{{\bf j},s}, c^\dag_{{\bf j}',s'} \} = \delta_{\bf j,j'} \delta_{s,s'}$, $n_{{\bf j},s} = c^\dag_{{\bf j},s} c_{{\bf j},s}$ 
is the density operator, $\mu_s$ the chemical potential, 
$U$ the interaction strength and $s=1,2$ labels the spin degree of freedom. 
The interaction Hamiltonian can be written also as
\beq
	H_{\rm int} = -\frac{2U}{3} \sum_{\bf j} {\bf S_j \cdot S_j},
	\label{eq H_int_S}
\eeq
where $S^a_{\bf j} = \frac{1}{2} c^\dag_{{\bf j},s} \sigma^a_{s,s'} c_{{\bf j},s'}$ are the usual spin operators.

To be more specific, when needed, we will consider the case of the gauge fields corresponding to the spin-orbit coupling of the Rashba type,
corresponding to the position independent tunnellings in the $x$ and $y$ directions:
\begin{subequations}
\label{soc}
 \begin{align} 
	T_x &= t{\rm e}^{-i \alpha \sigma_y} = 
t\left(
\begin{array}{cc}
 \cos (\alpha ) & - \sin (\alpha ) \\
 \sin (\alpha ) & \cos (\alpha )
\end{array}
\right) \\
	T_y &= t{\rm e}^{i \alpha \sigma_x} = 
t\left(
\begin{array}{cc}
 \cos (\alpha ) &  i\sin (\alpha ) \\
 i \sin (\alpha ) & \cos (\alpha )
\end{array}
\right).
 \end{align}
\end{subequations}
where $t$ denotes the global strength of the hopping amplitudes and we have used the labelling $T_x = T_{{\bf j,j}+1_x}$ and $T_y = T_{{\bf j,j}+1_y}$.


\subsection{Non-interacting case - $U=0$} 

In the case where the matrices $T_{\bf j,j'}$ are position independent, the Hamiltonian \eq{\ref{eq H parts kin}} +\eq{\ref{eq H parts mu}} is diagonalized in the momentum space: 
\beq
	H_0 = -\sum_{\bf k} C^\dag_{\bf k} \mathcal{T}_{\bf k} C_{\bf k},
	\label{eq H diag}
\eeq
where $C^{\dagger}_{\bf k} = (c^{\dagger}_{{\bf k},1}, c^{\dagger}_{{\bf k},2})$ and
\beqa
	\mathcal{T}_{\bf k} &=&  (T_x {\rm e}^{-ik_x} + T^\dag_x {\rm e}^{ik_x} +
 T_y {\rm e}^{-ik_y} + T^\dag_y {\rm e}^{ik_y} +\hat{\mu}) \nonumber \\
	\hat{\mu} &=& \left( \begin{array}{cc}
		 \mu_1  & 0 \\
		0  & \mu_2 \\
		\end{array} \right).
		\label{eq Tk}
\eeqa
 
Finally, the matrix $\mathcal{T}_{\bf k}$ is diagonalized with a unitary matrix $U_{\bf k}$, such that
\beq
H_0 = \sum_{{\bf k},s} \epsilon_{{\bf k},s} d^\dag_{{\bf k},s} d_{{\bf k},s}\label{eq H_0 diag},
\eeq
where $(d^{\dagger}_{{\bf k},1}, d^{\dagger}_{{\bf k},2})=(c^{\dagger}_{{\bf k},1}, c^{\dagger}_{{\bf k},2})U^\dag_{\bf k}$.

In the specific case of the spin-orbit coupling~\eqref{soc}, one obtains the following eigenenergies:
\beqa
	\epsilon_{{\bf k},1,2} &=& -\mu-2 t\biggl[ (\cos{k_x}+\cos{k_y}) \cos{\alpha} \biggr. \nonumber \\
	&& \biggl. \pm \sqrt{(\sin^2{k_x} + \sin^2{k_y}) \sin^2{\alpha} + h^2} \biggr],
	\label{eq eigenergies}
\eeqa
where $\mu$ is the chemical potential and $h$ is the spin imbalance defined through $\mu_{1,2} = \mu \pm h$ (see \eq{\ref{eq Tk}}).
The situation gets simpler in the limit of $\mu = h = 0$, which corresponds to half filling for all $\alpha$.

\subsection{Linear Response in small $U$ limit}

Turning on the interaction, one expects the Fermi liquid phase to be unstable towards a magnetically ordered phase. This instability of the system can be captured  using the standard linear response theory 
(see Appendix \ref{sec App susc} for details), more precisely from the spin-spin susceptibility. 
Let us start with the non interacting Hamiltonian of the form
\[
	H = H_0 + H_{\rm ext},
\]
where $H_0$ is given by \eq{\ref{eq H diag}} and
\[
	H_{\rm ext} = \sum_{\bf j} S_{\bf j}^a B_{\bf j}^a (t) = \frac{1}{N}\sum_{\bf q} S_{\bf -q}^a B_{\bf q}^a (t),
\]
where $B$ is the external driving force. Using the Fourier transform of the fermionic operators 
$c_{{\bf k},s} = \frac{1}{\sqrt{N}} \sum {\rm e}^{-i{\bf k \cdot j}} c_{{\bf j},s}$, 
where $N$ is the number of sites, one gets
 $S_{\bf j}^a = \frac{1}{N} \sum_{\bf q} {\rm e}^{i{\bf q \cdot j}} S^a_{\bf q}$ and 
$S^a_{\bf q} = \frac{1}{2}\sum_{\bf k} c^\dag_{{\bf k},s} \sigma^a_{s,s'} c_{{\bf k+q},s'}$. 
The spin-spin susceptibility is diagonal in the momentum space and reads
\beq
	\chi_{\bf q}^{a,b}(\omega) = -i \int_{0}^{+\infty} {\rm d}\tau\, {\rm e}^{-i\omega \tau}  
\left\langle \left[ S^a_{\bf q}(\tau),S^b_{-\bf q}(0) \right] \right\rangle,	
\eeq
where the thermal average and the time evolution are done using the unperturbed Hamiltonian $H_0$. Therefore, the analytic expression for
the susceptibility is easily obtained by diagonalizing $H_0$. 
After some manipulation, we find, that
\begin{widetext}
\beqa
		\chi_{\bf q}^{a,b}(\omega) &=& \frac{1}{N} \sum_{{\bf k},s,s'} 
(\mathcal{S}^a_{\bf k,k+q})_{s,s'} (\mathcal{S}^b_{\bf k+q,k})_{s',s} 
\frac{ n(\epsilon_{{\bf k},s'}) - n(\epsilon_{{\bf k+q},s})}{\omega + \epsilon_{{\bf k},s'} -\epsilon_{{\bf k+q},s} +i\eta}	\nonumber \\
		&\rightarrow& \frac{1}{(2\pi)^2} \int {\rm d}^2 k \sum_{s,s'} (\mathcal{S}^a_{\bf k,k+q})_{s,s'} (\mathcal{S}^b_{\bf k+q,k})_{s',s} F_{s,s'}(\omega,{\bf k,q}),
	\label{eq chi q_omega Main}
\eeqa
\end{widetext}
where $n(\epsilon) = (1+{\rm e}^{\beta \epsilon})^{-1}$ is the Fermi function. We have introduced
\beq
	F_{s,s'}(\omega,{\bf k,q}) = \frac{ n(\epsilon_{{\bf k},s'}) - n(\epsilon_{{\bf k+q},s})}{\omega +
 \epsilon_{{\bf k},s'} -\epsilon_{{\bf k+q},s} +i\eta}
	\label{eq integrand F}
\eeq
and 
\beq
	\mathcal{S}^a_{\bf k,k+q} = \frac{1}{2} U_{\bf k} \sigma^a U^\dag_{\bf k+q},
	\label{eq S def}
\eeq
where $\eta$ is an infinitesimal convergence factor. In the usual situation of a Hamiltonian diagonal in the original spin-space, the matrices
$U_{\mathbf{k}}$ are simply the identity and one recovers the standard spin-spin susceptibility.  \\

In what precedes, we have derived the susceptibility \eq{\ref{eq chi q_omega Main}} for the non-interacting system. However, 
the interaction among the fermions affects the spin-spin susceptibility. This can be captured, in the random phase approximation (RPA) 
framework, by deriving, in a self-consistent way, the effective propagator for the spin fluctuations. This is equivalent to perform a mean-field
approximation to the interacting Hamiltonian~\cite{Jensen_1991,Altland_n_Simons_2010,Demler11}.

 Lets recall the main step: Starting from the interaction Hamiltonian \eq{\ref{eq H_int_S}}, 
the effect of interaction then amounts to an introduction of an effective driving force given by
\beq
	H^{\rm eff}_{\rm ext} = H_{\rm ext} + H^{\rm MF}_{\rm int} = \sum_{\bf q} S^b_{\bf -q}(t) (B^b_{\bf q})^{\rm eff}(t),
\eeq
where $(B^b_{\bf q})^{\rm eff}(t) = \left[ B^b_{\bf q}(t) - 2g\langle S^b_{\bf q}\rangle(t) \right]$ and 
we have introduced $g=2U/3$ (see Appendix \ref{sec App susc}). 
After the manipulation described in Appendix \ref{sec App susc} one finds the expression for average values of the spin operators
\beq
	\langle{\mathbf{S}}_{\bf q}\rangle(\omega) = M^{-1}_{\bf q}(\omega) \chi_{\bf q}(\omega) {\bf B_{\bf q}}(\omega)
\label{rpachi}
\eeq
where $M^{ab}_{\bf q}(\omega)=\delta^{a,b} + 2g \chi^{a,b}_{\bf q}(\omega)$. $M^{-1}_{\bf q}(\omega) \chi_{\bf q}(\omega)$
is therefore the RPA susceptibility, whose singularities in the complex $\omega$ plane correspond to the
vanishing eigenvalues of $M_{\bf q}(\omega)$.\\

\subsection{Large $U$ limit - the Heisenberg Hamiltonian}

Following the method of~\cite{MacDonald_1988}, one obtains,  in the large (repulsive) interaction $U$ limit and at
half-filling, the following effective Heisenberg Hamiltonian, up to the second order in the $t/U$ expansion: 

\begin{widetext}
\beq
 H = H_c + \sum_{\delta=x,y} \sum_{<i,i+\delta>} \sum_{a=x,y,z} J^a_\delta S^a_i S^a_{i+\delta} + \mathbf{D}_{\delta +} 
\cdot (\mathbf{S}_i \times \mathbf{S}_{i+\delta})_+ + \mathbf{D}_{\delta -} \cdot (\mathbf{S}_i \times \mathbf{S}_{i+\delta})_-,
 \label{eq H Heis gen}
\eeq
\end{widetext}
where
\beqa
	(\mathbf{S}_i \times \mathbf{S}_{i+\delta})_+ &=& (S^y_1 S^z_2, S^z_1 S^x_2, S^x_1 S^y_2) \nonumber \\
	(\mathbf{S}_i \times \mathbf{S}_{i+\delta})_- &=& -(S^z_1 S^y_2, S^x_1 S^z_2, S^y_1 S^x_2) \nonumber
\eeqa
are the "positive" and "negative" part of the vector product. In the most general case, 
the coefficients $J^a_\delta, D^p_\delta$ and $H_c$, $\delta=x,y$, $a=x,y,z$, $p=1,2,3$, are quadratic 
functions of elements of the tunnelling matrices $T$. The general expression can be found in the appendix~\ref{appheis}. However, 
the situation will simplify considerably when considering a specific case of spin orbit coupling of Rashba type, see Eq.~\eqref{soc}:

\beqa
 J^x_\delta &=& 4 \lambda \cos{(2 \alpha)} \nonumber \\
 J^y_\delta &=& 4 \lambda  \nonumber \\
 J^z_\delta &=& 4 \lambda \cos{(2 \alpha)} \nonumber
\eeqa
for both spatial directions $\delta=x,y$,
\beqa
	D^x_{\delta +} &=& D^x_{\delta -} = 0 \nonumber \\
	D^y_{\delta +} &=& D^y_{\delta -} = -4 \lambda \sin{(2 \alpha )}  \nonumber \\
	D^z_{\delta +} &=& D^z_{\delta -} = 0 \nonumber   
\eeqa

for $\delta=x$ (tunnelling $T_x$) and 

\beqa
	D^x_{\delta +} &=& D^x_{\delta -} = -4 \lambda \sin{(2 \alpha )} \nonumber \\
	D^y_{\delta +} &=& D^y_{\delta -} = 0  \nonumber \\
	D^z_{\delta +} &=& D^z_{\delta -} = 0 \nonumber   
\eeqa

for $\delta=y$, (the tunnelling $T_y$),

\[
	H_c =  -4 \lambda \frac{\mathds{1}}{4},
\]

$\lambda = t^2/U$. This Hamiltonian is identical with Eq. (2) of \cite{Cole_2012}, where one has to set their parameter $\lambda=1$. 
In the following, we take $t$ equal to unity to set the energy scale.


\section{Results \& Discussion}
\label{result}

\subsection{Small U limit}

\begin{center}
	\begin{figure}[t!p]
  	\includegraphics[width=9cm]{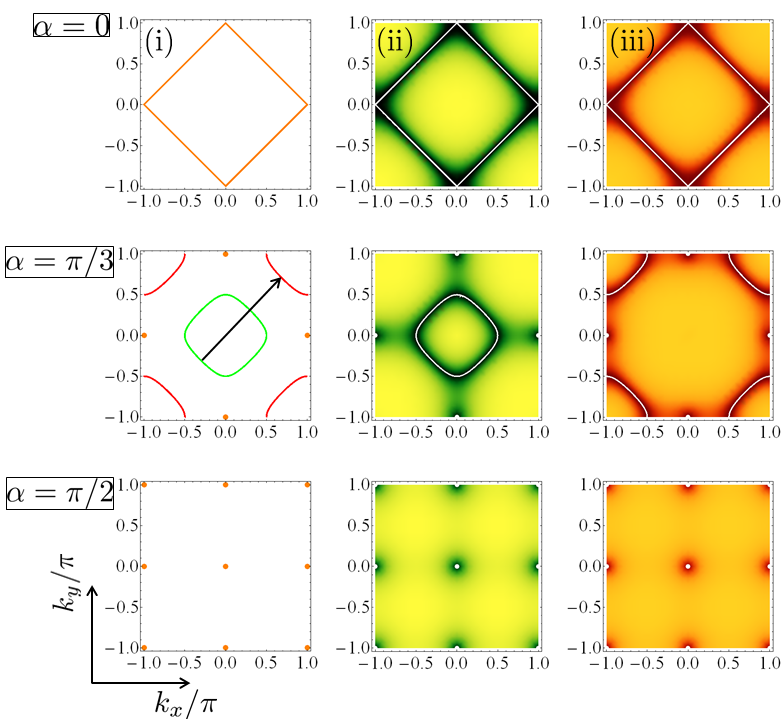} 
  	\caption{(Color online) Plots of Fermi surfaces and susceptibility integrand $F$ (see \eq{\ref{eq integrand F}}) in the first BZ for ${\bf q} = (\pi, \pi)$. Rows: 
  	successive values of $\alpha=0, \pi/3, \pi/2$; Columns: (i) Fermi surfaces. 
  	The nesting of Fermi surfaces is indicated explicitly for $\alpha=\pi/3$. 
  	For $\alpha=\pi/2$ the Fermi surface becomes a set of isolated Dirac points. 
  	(ii) $F$ for spins ($s,s’$)=(2,1), (iii) $F$ for spins ($s,s’$)=(1,2). 
  	The green and red colours in (ii) and (iii) represent the maximum of $F$. 
  	The Fermi surfaces for given $(s,s')$ in (ii) and (iii) are indicated by white lines. 
  	The coincidence of Fermi surfaces and the maximum of the integrand $F$ is not generic, 
  	but is specific for ${\bf q}_\pi = (\pm \pi, \pm \pi)$. } 
  	\label{fig Fermi surf}
 	\end{figure}
\end{center}

Here, we study the small interaction behaviour of the lattice gas at half-filling, $\mu=h=0$, 
by means of the susceptibility, given by \eq{\ref{eq chi q_omega Main}}. 
The instability of the non-interacting ground state is signalled by a divergence of the DC ($\omega=0$)
RPA susceptibility~\eqref{rpachi}, corresponding to a vanishing eigenvalue of $M_{\mathbf{q}}(0)$. 
The latter corresponds to an eigenvalue of $\chi_{\mathbf{q}}(0)$ having the value $-1/2g$.
In the following, we evaluate the susceptibility by numerical integration of \eqref{eq chi q_omega Main} 
over the first Brillouin zone. We will discuss the general $\mathbf{q}$ value below, but
it turns out that the mean-field simulations indicate the onset of AF phase for small couplings $U$,
 corresponding to a value of ${\bf q}={\bf q}_\pi = (\pm \pi,\pm \pi)$.
For this specific value of ${\bf q = q}_\pi$, one can see, 
using \eq{\ref{eq eigenergies}}, that $\epsilon_{{\bf k +q}_\pi,s} = -\epsilon_{{\bf k},\bar{s}}$, 
where $\bar{s}$ means the opposite spin to $s$.  Moreover, in this case, the Fermi energy is $E_F=0$ and 
one gets nested Fermi surfaces for any value of the gauge field parameter $\alpha$ - 
the Fermi surfaces together with the susceptibility integrand function 
$F_{s,s'}(0,{\bf k},(\pi,\pi))$, \eq{\ref{eq integrand F}}, are plotted in \fig{\ref{fig Fermi surf}} for different values of $\alpha$. 
The coincidence of the maximum of the integrand function $F$ is a specific feature of ${\bf q}_\pi$ and 
is responsible for the the onset of AF phase for sufficiently low temperature, i.e. large $\beta$ values. Indeed, we note, 
that in the case of nested Fermi surfaces, the susceptibility possess ${\rm ln} \beta$ 
divergence. 
Moreover, for $\alpha=0$, the presence of Van Hove singularities at the Fermi energy 
adds a leading divergence ${\rm ln}^2\beta$. Therefore, 
we fit the susceptibility with a function
\beq
	\chi_{\rm fit}= a \ln^2{\beta} + b \ln{\beta} + c.
	\label{eq fitting fction}
\eeq
{\col{The results are summarized in Table \ref{tab fit} and plotted in \fig{\ref{fig Susc vs. beta}}. With respect to the preceding discussion, one can distinguish two cases - $\alpha=0$ and $\alpha>0$. We verified, that for $\alpha=0$, our results agree with the theoretical prediction given by Eq.(7) in~\cite{Shimahara_1988, Note1} (first two lines in Table \ref{tab fit})
\beq
	\chi_{{\bf q}_\pi,{\rm theor}} = -\frac{1}{2\pi^2 \tilde{t}} \ln^2{\frac{16 {\rm e}^\gamma \tilde{t}}{\pi} \beta} + C_0.
	\label{eq chi theor}
\eeq
For $\alpha>0$ ($\alpha$ strictly positive), the absolute value of $b$ decreases monotonically as the nesting of Fermi surfaces decreases. Also, for $\alpha>0$ the $a$ term becomes significantly smaller than for the $\alpha=0$ case (no $\ln^2 \beta$ divergence).}}

\begin{table}
\begin{tabular}{|c|c|c|c|c|}
\hline
 $\alpha$  & $a\times 10^3$ & $b$ & $c$ & $\sigma^2\times 10^6$ \\ \hline
 0 (theor) & -13. & -0.073 & -0.098 & 120. \\ \hline
 0 (fit) & -13. & -0.073 & -0.1 & 0.73 \\ \hline
 0.05 & 3. & -0.13 & -0.057 & 6.6 \\ \hline
 $\pi/12$ & 0.9 & -0.065 & -0.11 & 2. \\ \hline
 $\pi/6$ & 0.38 & -0.043 & -0.1 & 2.7 \\ \hline
 $\pi/4$ & -0.76 & -0.025 & -0.1 & 1.6 \\ \hline
 $\pi/3$ & -0.68 & -0.016 & -0.088 & 0.032 \\ \hline
 $5\pi/12$ & -0.85 & -0.0052 & -0.082 & 1.6 \\ \hline
 $\pi/2$ & -0.95 & 0.0047 & -0.084 & 1.4\\
 \hline 
\end{tabular}
 \caption{Values of the fitting parameters $a,b,c$ of \eq{\ref{eq fitting fction}} for 
different values of the gauge field parameter $\alpha$. $\sigma^2$ is the usual data variance, defined in the text. {\col{The first line corresponds to the theoretical prediction \eq{\ref{eq chi theor}} and agrees well with the numerical computation of the susceptibility \eq{\ref{eq chi q_omega Main}} shown in the second line. }}
}
 \label{tab fit}
\end{table}
\begin{center}
	\begin{figure}[h!]
  	\includegraphics[width=8cm]{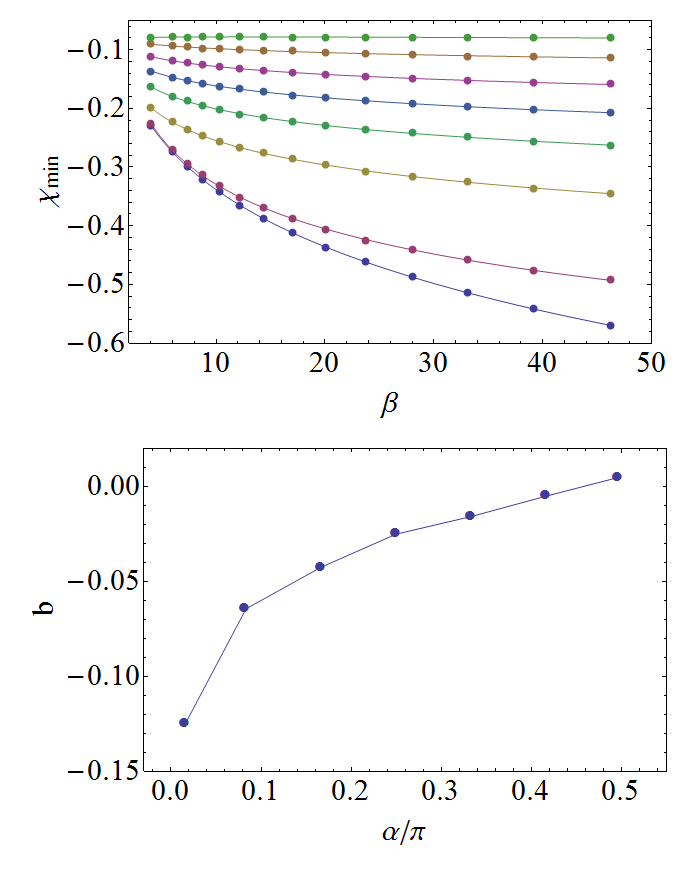} 
  	\caption{(Color online) a) Plot of the minimal eigenvalue of $\chi$ against $\beta$ for ${\bf q}=(\pi,\pi)$. 
  	The data points were calculated using \eq{\ref{eq chi q_omega Main}} and the lines are the fits \eq{\ref{eq fitting fction}}. 
  	The lowest curve corresponds to $\alpha=0$ and the highest to $\alpha=\pi/2$ 
  	in monotonically increasing order corresponding to Table \ref{tab fit}. 
  	b) {\col{Fitting parameter $b$ against the gauge field parameter $\alpha>0$ (only non zero values of $\alpha$). The plot shows a monotonic decrease 
  	of $|b|$ - see text for details. }} } 
  	\label{fig Susc vs. beta}
  	\setlength{\unitlength}{1cm}	
  	\begin{picture}(2,2)
		  \put(-3.2,14.7){\Large{ a) }}
		  \put(-3.2,9.8){\Large{ b) }}
		\end{picture}
 	\end{figure}
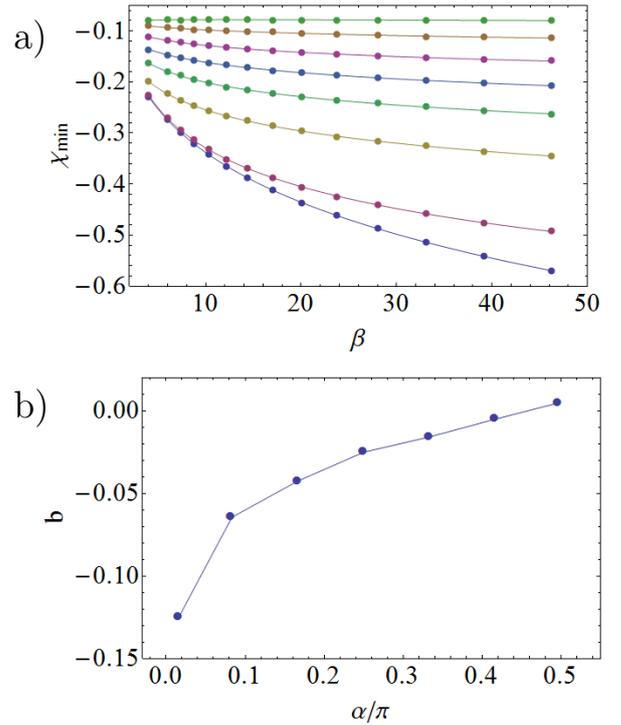
\end{center}
The data variance in the Table \ref{tab fit} reads $\sigma^2 = \frac{1}{n}\sum_{i=1}^n |\chi_{{\rm fit},i} - \chi_i|^2$, 
where $\chi_i$ stands for a minimal eigenvalue at (given) ${\bf q}_\pi$ and $n=14$ is the number of simulated data points. 
As shown in \fig{\ref{fig Susc vs. beta}}, the parameter of the ${\rm ln} \beta$ divergence $b$ increases 
monotonically (for $\alpha>0$) with $\beta$. This comes from the fact, that the contribution to 
the susceptibility \eq{\ref{eq chi q_omega Main}} is proportional to the area of the Fermi surface 
(length of the 1D surface in our case), which decreases with increasing $\alpha$ and shrinks 
to the Dirac points for $\alpha = \pi/2$ as shown in \fig{\ref{fig Fermi surf}}.

\begin{center}
	\begin{figure}[t!]
  	\includegraphics[width=8cm]{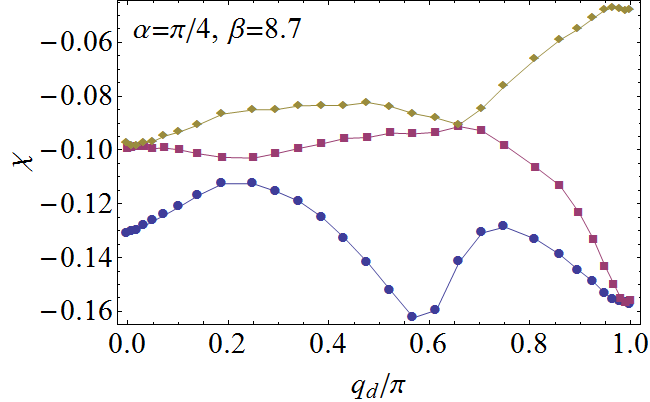} 
  	\caption{(Color online) Susceptibility eigenvalues vs. $\bf q$ on the diagonal of the Brillouin zone, ${\bf q} = (q_d,q_d)$. An example for $\alpha=\pi/4$ and $\beta=8.7$ is shown and demonstrates a minimum of the susceptibility occurring for ${\bf q} \neq {\bf q}_\pi$. } 
  	\label{fig Susc vs. q}
 	\end{figure}
\end{center}

Even though the AF order is expected at $T=0$, at higher temperature the susceptibility develops minima at ${\bf q} \neq {\bf q}_\pi$. 
An example for a diagonal ${\bf q}$, ${\bf q} = (q_d, q_d)$, $q_d \in \left[0, \pi \right]$ is 
shown in \fig{\ref{fig Susc vs. q}}.  Therefore, at this temperature, the linear response analysis predicts 
an instability towards a different magnetic order. In addition, since these minima correspond
to a finite value of the susceptibility $\chi_m$, the phase transition is predicted  to occur at a finite interaction $U=-\frac{3}{4}\frac{1}{\chi_m}$. 
However, this prediction assumes a second order phase transition and at this (large) value of $U$, another magnetic order might have already 
appeared. In addition, since the value of the minimum of the susceptibility is not much lower than the one of the AF order, it might explain 
that, from a numerical point of view (finite size...), the onset of those non-AF phases at finite temperature and finite interaction 
have not been observed yet. 

~\\
~\\
~\\

\subsection{Mean Field numerical simulation}
To further investigate the properties of the system, we study the properties of the mean-field Hamiltonian ground state. 
More precisely, at finite temperature, we minimize the mean-field free energy $F_{\mathrm{MF}}=-\frac{1}{\beta}\ln{Z}$, where $Z$ is the
partition function associated to the mean-field Hamiltonian $H_{\mathrm{MF}}$ (See Appendix~\ref{sec App susc} for details):
\begin{equation}
\label{meanfieldh}
\begin{aligned}
H_{\mathrm{MF}}= &- \sum_{\bf j,j'} c^\dag_{{\bf j},s} T^{s, s'}_{\bf j, j'}  c_{{\bf j'},s'} -\frac{4U}{3} \sum_{\bf j} {\langle\mathbf{S_j}\rangle \cdot \mathbf{S_j}} \\
	 &+\frac{2U}{3}\sum_{\bf j} {\langle\mathbf{S_j}\rangle \cdot \langle \mathbf{S_j}\rangle}.
\end{aligned}
\end{equation}
At half-filling, with a repulsive interaction, the total average density is 
expected to remain fixed to unity, the relevant degrees of freedom being the average values of the spin operators $S_j^a$. 
The present mean-field decoupling respects thus the $SU(2)$ invariance of the interaction and allows for all possible magnetic
orderings, in particular those obtained in the large $U$ limit from the effective Heisenberg model~\cite{Cole_2012,Hofstetter12}. From that point of view,
even though other mean-field decoupling schemes are possible~\cite{schultz90}, the present one is expected to capture 
qualitatively the properties of the magnetic phases for different values of $U$ and $\alpha$.

The numerical calculation has been performed on a $36 \times 36$ square lattice with periodic boundary conditions
for different values of parameters $\alpha, U, \beta$. On each lattice site, the three components of the spin $\langle\mathbf{S_j}\rangle$ are 
independent mean-field parameters. Since the mean-field Hamiltonian~\eqref{meanfieldh} is quadratic in the fermionic operators, the free
energy can be obtained by diagonalizing a $2N\times 2N$ matrix, where $N$ is the number of lattice sites. Even though the exact 
structure of the matrix is slightly different from the one obtained in the BCS case~\cite{Dubi_07,Chen09,Iskin13}, there is a one to one mapping between the
two situations, namely a particle-hole transformation on one of the species. From a numerical point of view, the free energy is minimized
using a mixed quasi-Newton and conjugate gradient method; additional checks (e.g. different initial values of the spins) 
were performed to ensure that the global minimum has been reached. Finally, even though the mean-field calculation captures the temperature
dependence of the spin degrees of freedom, it only  describes the Mott transition, whereas the determination of the true critical temperature 
to a quasi-long range magnetic order, especially in the large interaction limit, 
amounts to taking into account the effects of terms beyond mean field~\cite{Sademelo97}.

The results are summarized in Table~\ref{tab Summary}. 
\begin{table}
\begin{tabular}{c}
  	\includegraphics[width=8cm]{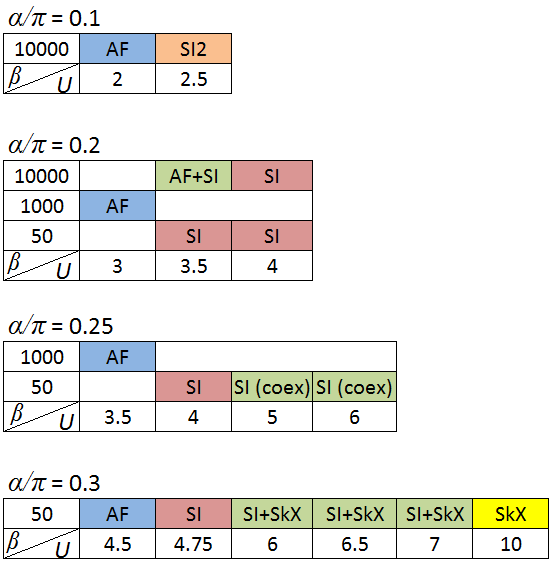}
\end{tabular}  	
  	\caption{ (Color online) Summary of the MF results. \emph{Legend:} AF (blue) - antiferromagnetic phase, SI (dark red), SI2 (orange) - spiral phases, SkX (yellow) - Skyrmion crystal, see text and Table \ref{tab phase orderings} for definitions; For fixed values of $\alpha$, we show phases which occur as a function of $\beta$, the inverse temperature, and $U$, the interaction strength. For small values of the interaction, AF phase occurs for all $\alpha$ and for sufficiently high values of $\beta$ (low temperatures), in agreement with the results predicted by the linear response theory. For increasing values of $U$, a phase transition occurs towards different phases (SI, SI2, SkX), depending on $\alpha$. For other values of $\alpha$, not shown here, we have found a similar scenario. For $\alpha\geq0.4\pi$, the logarithmic divergence of the susceptibility with the temperature is very slow, such that the AF phase only appears for very low values of the temperature and, from a numerical point of view, is very difficult to observe with our method. Coexistence of different phases is indicated (either of two well defined phases or of a well defined phase and an unknown phase (coex)).}
  	\label{tab Summary}
\end{table}
Only few values of the gauge field and interaction are presented, since the 
focus of the paper is on the generic evolution of the system phase from the non-interacting situation to the large interacting limit.  For the parameters under consideration, we have identified the following phases: • Antiferromagnet (AF), ${\bf q}=(\pi, \pi)$, • Spiral (SI), ${\bf q} = (q, q')$, • Spiral 2 (SI2), ${\bf q} = (q, \pi)$, •Skyrmion crystal (SkX), ${\bf q} = (\pi, \pm \pi/3)$  and ${\bf q}$ is modulo the periodicity of the BZ. 
As one can see from the Table~\ref{tab Summary}, an AF phase always shows up first at low interaction; one can see that when the transition normal-AF occurs 
for a low interaction, the phase only exists for very low temperature, a behaviour similar to the BCS situation.  
{\col{For other values of $\alpha$, not shown here, we have found a similar scenario. 
For $\alpha\geq0.4\pi$, the logarithmic divergence of the susceptibility with the temperature is very slow, 
 such that the AF phase only appears for very low values of the temperature and, from a numerical point of view, is very difficult to observe 
with our method.}}

\begin{figure*}[t!]	
  	\includegraphics[width=17cm]{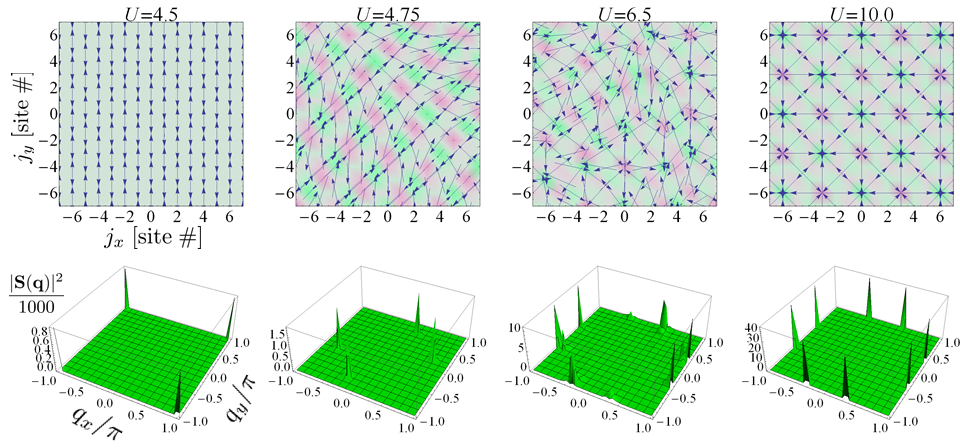}
  	\caption{(Color online) \emph{First row:} Density vector plots of spins $\braket{S^a_{\bf j}}$ for different interaction strengths $U=4.5, 4.75, 6.5, 10$. We plot the central region of the $36\times36$ lattice. The real space $z$ component of the spin $S_{\bf j}^z$ is color encoded, where the pink color corresponds to the maxima and the green color to the minima of $S^z$ respectively. The color scaling is relative to each $U$ independently. \emph{Second row:} 3D plots of the spin density $|S({\bf q})|^2$ in the first Brillouin zone for interaction strengths $U=4.5, 4.75, 6.5, 10$: $U = 4.5, 4.75, 10$ correspond to AF, SI and SkX phases, which are clearly indicated by peaks of $|S({\bf q})|^2$. $U = 6.5$ corresponds to a crossover between SI and SkX phase, where the two phases coexist. The used parameter values are $\alpha = 0.3\pi$ and $\beta=50$. See also Table \ref{tab Summary} and Appendix \ref{sec App mag orders} for quantitative details.} 
  	\label{fig Sk sq}
\end{figure*}

For larger interaction values, the AF order further evolves to a phase depicting a magnetic texture (spiral...), corresponding to peaks in 
$|\mathbf{S(q)}|^2$ away from the corner of the BZ. The transition is of first order, since the AF order parameter doesn't vanish at the transition point, at which also a magnetic texture appears. 

Then, for larger interaction strength and depending on the gauge-field parameter, the magnetic texture can further evolve to another phase, 
 to reach finally the spin configuration predicted from the Heisenberg-model. It's more difficult to determine the type of transition 
from one phase to the other, but from the numerical data it seems to correspond to a smooth change in the location of the peaks of
$|\mathbf{S(q)}|^2$, i.e. a crossover within the numerical resolution of the simulation.

Finally, one should notice that the critical value $U_c(\alpha)$ of the transition from the AF order is not expected to depict a smooth
curve in the $U - \alpha$ plane, since, in the magnetic texture phase, the different mean-field Hamiltonians $H_{\mathrm{MF}}(\alpha)$ have 
different translational symmetries, a situation similar to the Hofstadter-Hubbard model, depicting fermions (or bosons) in 
an external magnetic field. Nevertheless, our numerics seems to suggest that the transition from an AF order to a magnetic texture phase
increases with the gauge-field parameter $\alpha$, such that the AF order on the $\alpha=0$ line seems to be 
unstable towards a magnetic structure phase in the presence of arbitrary small spin-orbit coupling.

In \fig{\ref{fig Sk sq}}, an example of a real space spin configuration $\braket{S^a_{\bf j}}$ 
together with its Fourier transform $|S({\bf q})|^2$ is shown for $\alpha = 0.3\pi$ and $\beta=50$. 
Different phases (AF, SI, SkX for $U$=4.5, 4.75, 10) or their coexistence (SI + SkX for $U$ = 6.5) 
are clearly indicated by peaks of $|S({\bf q})|^2$. As one can see, in the large $U$ limit, one recovers the spin 
configuration expected from the effective Heisenberg model, i.e. a $3\times3$ Skyrmion crystal which is non-planar
magnetic order with a non-vanishing Skyrmion density $\mathbf{S_j}\cdot\left(\mathbf{S_{j+1_x}}\times\mathbf{S_{j+1_y}}\right)$. The magnetic orders can be parametrized by the peak values of $|{\bf S(q)}|^2$. More specifically, each peak ${\bf q}$ gives rise to a spin wave, which can be described as \cite{Sachdev_2003}
\beq
 \braket{S^a_{\bf j}} = N_1^a({\bf q}) \cos{\bf q \cdot j} + N_2^a({\bf q}) \sin{\bf q \cdot j}
 \label{eq mag order}
\eeq
with further distinction of collinear, ${\bf N}_1 ({\bf q}) \times {\bf N}_2 ({\bf q}) = 0$, and 
non collinear, ${\bf N}_1 ({\bf q}) \times {\bf N}_2 ({\bf q}) \neq 0$, orders. 
The values of ${\bf q}, {\bf N}_1({\bf q})$ and ${\bf N}_2({\bf q})$ are summarized in Appendix \ref{sec App mag orders}.

\begin{table}
 \begin{tabular}{|c|c|c|c|}
 \hline
  & ${\bf q}$ & ${\bf N}_1({\bf q}) \times {\bf N}_2({\bf q})$ & ${\bf N}_1({\bf q}) \cdot {\bf N}_2({\bf q})$ \\ \hline
  AF & $(\pi, \pi)$ & 0 & 0 \\ \hline
  SI & $(q, q')$ & $\neq$ 0 & 0 \\ \hline
  SI2 & $(q, \pi)$ & $\neq 0$ & 0 \\ \hline
  SkX & $(\pi, \pm \pi/3)$ & non-planar & $\neq 0$ \\
  \hline 
 \end{tabular} 
 \caption{Summary of magnetic phases. Different phases are defined by the value of ${\bf q}$ at which $|{\bf S(q)}|^2$ peaks. 
 Further distinction of magnetic orders given by the values of ${\bf N_1 \cdot N_2}$ and ${\bf N_1 \times N_2}$ is shown. 
 The skyrmion phase (SkX) corresponds to a pair of parameters $({\bf N}_1({\bf q_{\pm}}),{\bf N}_2({\bf q_{\pm}}))$ at inequivalent positions 
 ${\bf q_{\pm}}$ in the Brillouin zone, with, in addition, non-collinear ${\bf N}_1({\bf q_{\pm}}) \times {\bf N}_2({\bf q_{\pm}})$ vectors.
 See also \eq{\ref{eq mag order}} and the text for details. }
 \label{tab phase orderings}
\end{table}

\section{Conclusion}

In summary, we have studied the quantum phase transitions of the Fermi-Hubbard model in a square lattice at half-filling in the presence of an effective
spin-orbit coupling. We have shown that at small interaction, the system always enters an AF order, then undergoes a first order 
phase transition to a  phase  depicting magnetic texture, and, eventually, reaches (at large interaction strength), the magnetic texture 
predicted by the associated Heisenberg model.

In addition to the half-filling situation presented here, a possible study will concern the doped case or the imbalanced population case, where
more exotic magnetic phases are expected to occur, possibly in competition with the non-conventional superconductivity. 
One should also take into account the effects of terms beyond mean field to determine properly the
critical temperature of the transition and to estimate the
strength of the quantum fluctuations thus allowing for a better
comparison with possible experimental results.

Apart from the (magnetic) properties of the ground state, it would be interesting to study the excitations above the ground state,
in particular to describe the dynamical response of the system to external perturbations, like the (sudden) quenches of 
the interaction or the gauge-field, which can efficiently be achieved in cold
atomic gases experiments.


\newpage

\appendix

\begin{widetext}


\section{Linear response of the spin systems}
\label{sec App susc}
 From the linear response theory \cite{Jensen_1991, Altland_n_Simons_2010}, the expression for the spin-spin susceptibility $\chi_{\bf q}^{a,b}(\tau)$ reads:
\beq	
		\chi_{\bf q}^{a,b}(\tau) = -i\theta(\tau)\left\langle \left[ S^a_{\bf q}(\tau),S^b_{\bf -q}(0) \right] \right\rangle.
	\label{eq chi q_omega def}
\eeq
The frequency domain susceptibility is given by the Fourier transform $\chi_{\bf q}^{a,b}(\omega)=\int d\tau\,e^{-i\omega \tau}\chi_{\bf q}^{a,b}(\tau)$.
~\\

We can now find an explicit analytical expression 
for $\chi^{a,b}_{\bf q}(\omega)$, given the Hamiltonian \eq{\ref{eq H parts kin} - \ref{eq H parts int}}. 
Namely we need to evaluate the thermal average
\beq
	\left\langle \left[ S^a_{\bf q}(\tau),S^b_{\bf -q}(0) \right] \right\rangle = 
Z^{-1} {\rm Tr} \biggl[ \left(S^a_{\bf q}(\tau) S^b_{\bf -q}(0) - S^b_{\bf -q}(0) S^a_{\bf q}(\tau)\right) {\rm e}^{-\beta H_0} \biggr].
	\label{eq Tr comm}
\eeq
 In order to evaluate the trace and the time dependence, one needs to diagonalize $H_0$. As explained in the main text,
this is achieved by going to momentum space and finding $2\times2$ unitary transform $U_{\mathbf{k}}$ ($d_{\bf k} = U_{\bf k} c_{\bf k}$) such that 
\beq
	H_0 = \sum_{{\bf k},s} \epsilon_{{\bf k},s} d^\dag_{{\bf k},s} d_{{\bf k},s}.
\eeq
In the diagonal basis, the time evolution of the operators $d_{{\bf k},s}$ is simple, such that one obtains readily the
time evolution of the spin operators:
\beq
	S^a_{\bf q}(\tau) = \sum_{\bf k} d^\dag_{{\bf k},s} (\mathcal{S}^a_{\bf k,k+q})_{s,s'} 
d_{{\bf k+q},s'} {\rm e}^{i\tau(\epsilon_{{\bf k},s} - \epsilon_{{\bf k+q},s'} )},
	\label{eq St}
\eeq
where
\beq
	\mathcal{S}^a_{\bf k,k+q} = \frac{1}{2} U_{\bf k} \sigma^a U^\dag_{\bf k+q}.
\eeq
We proceed with the evaluation of the trace \eq{\ref{eq Tr comm}}. Lets start with the first part of the commutator
\beqa
	{\rm Tr} \left( S^a_{\bf q}(\tau) S^b_{\bf -q}(0) {\rm e}^{-\beta H_0} \right) &=& 
\sum_{n_i} \bra{n_1} S^a_{\bf q}(\tau) \ket{n_2}\bra{n_2} S^b_{\bf -q}(0) \ket{n_3} \bra{n_3} {\rm e}^{-\beta H_0} \ket{n_1} \nonumber \\
	&=&  \sum S^a_{\bf q}(\tau)_{n_1,n_2} S^b_{\bf -q}(0)_{n_2,n_1}  \left[ {\rm e}^{-\beta H_0} \right]_{n_1,n_1},
	\label{eq Tr of SS}
\eeqa
where the sum runs over complete basis (i.e. momentum and spin) 
$\ket{n_i}$ and we have used the fact, that $H_0$ is already diagonalized.
 Plugging the expression \eq{\ref{eq St}} to \eq{\ref{eq Tr of SS}} one finds
\beqa
	&&{\rm Tr} \left( S^a_{\bf q}(\tau) S^b_{\bf -q}(0) {\rm e}^{-\beta H_0} \right) = \nonumber \\
	&=& {\rm Tr} \left( 
	d^\dag_{{\bf k},s} (\mathcal{S}^a_{\bf k,k+q})_{s,\sigma} 
d_{{\bf k+q},\sigma} d^\dag_{{\bf k'},s'} (\mathcal{S}^b_{\bf k',k'-q})_{s',\sigma'} d_{{\bf k'-q},\sigma'} 
{\rm e}^{-i\tau(\epsilon_{{\bf k'},s'} - \epsilon_{{\bf k'-q},\sigma'} )}  
{\rm e}^{-\beta \sum \epsilon_{{\bf p},\sigma} d^\dag_{{\bf p},\sigma} d_{{\bf p},\sigma} }
	\right) \nonumber \\
	&=& {\rm Tr} \left( 
	d^\dag_{{\bf k},s} (\mathcal{S}^a_{\bf k,k+q})_{s,s'} d_{{\bf k+q},s'} 
d^\dag_{{\bf k+q},s'} (\mathcal{S}^b_{\bf k+q,k})_{s',s} d_{{\bf k},s} 
{\rm e}^{-i\tau(\epsilon_{{\bf k+q},s'} - \epsilon_{{\bf k},s} )}  {\rm e}^{-\beta \sum \epsilon_{{\bf p},\sigma} 
d^\dag_{{\bf p},\sigma} d_{{\bf p},\sigma} }
	\right).
	\label{eq Tr SqSmq}
\eeqa
Following \eq{\ref{eq Tr of SS}}, we put in the previous ${\bf k' = k+q}$ and $\sigma = s', \; \sigma' = s$. One obtains
\beqa
	{\rm Tr} && \left( S^a_{\bf q}(\tau) S^b_{\bf -q}(0) {\rm e}^{-\beta H_0} \right) = \nonumber \\
	&=& Z\, \sum_{{\bf k},s,s'} (\mathcal{S}^a_{\bf k,k+q})_{s,s'} (\mathcal{S}^b_{\bf k+q,k})_{s',s} 
{\rm e}^{-i\tau(\epsilon_{{\bf k+q},s'}-\epsilon_{{\bf k},s})} \frac{ {\rm e}^{-\beta \epsilon_{{\bf k},s}} }{(1+{\rm e}^{-\beta \epsilon_{{\bf k},s}})(1+{\rm e}^{-\beta \epsilon_{{\bf k+q},s'}})} , \;\; {\bf q} \neq 0   \nonumber \\
	&=& Z\, \left\{ 
	\sum_{{\bf k},s \neq s'} (\mathcal{S}^a_{\bf k,k})_{s,s'} (\mathcal{S}^b_{\bf k,k})_{s',s} 
{\rm e}^{-i\tau(\epsilon_{{\bf k},s'}-\epsilon_{{\bf k},s})} \frac{ {\rm e}^{-\beta \epsilon_{{\bf k},s}} }{(1+{\rm e}^{-\beta \epsilon_{{\bf k},s}})(1+{\rm e}^{-\beta \epsilon_{{\bf k},s'}})}  \right.  \nonumber \\
	&& \left.  + \sum_{{\bf k},s} (\mathcal{S}^a_{\bf k,k})_{s,s} (\mathcal{S}^b_{\bf k,k})_{s,s} 
{\rm e}^{-i\tau(\epsilon_{{\bf k},s'}-\epsilon_{{\bf k},s})} \frac{ {\rm e}^{-\beta \epsilon_{{\bf k},s}} }{1+{\rm e}^{-\beta \epsilon_{{\bf k},s}}} 
	\right\}, \;\; {\bf q} = 0.
	\label{eq Tr eval}
\eeqa
The trace of $S^b_{\bf -q}(0) S^a_{\bf q}(\tau) {\rm e}^{-\beta H_0}$ is obtained in a similar way and yields a result identical to \eq{\ref{eq Tr eval}} with exchange $\beta \epsilon_{{\bf k},s} \leftrightarrow \beta \epsilon_{{\bf k+q},s'}$ (i.e. the energies are exchanged only in the thermal terms including $\beta$). We next notice, that
\beqa
	\frac{{\rm e}^{-\beta\epsilon_a} - {\rm e}^{-\beta\epsilon_b}}{(1+{\rm e}^{-\beta\epsilon_a})(1+{\rm e}^{-\beta\epsilon_b})} &=&  \frac{{\rm e}^{-\beta\epsilon_a} + 1 - 1 - {\rm e}^{-\beta\epsilon_b}}{(1+{\rm e}^{-\beta\epsilon_a})(1+{\rm e}^{-\beta\epsilon_b})} \nonumber \\
	&=& \frac{1}{1+{\rm e}^{-\beta\epsilon_b}} - \frac{1}{1+{\rm e}^{-\beta\epsilon_a}} = (1-n(\epsilon_b)) - (1-n(\epsilon_a)) \nonumber \\
	&=& n(\epsilon_a) - n(\epsilon_b).
\eeqa
This also holds for the degenerate case ($\bf q=0$, last line in \eq{\ref{eq Tr eval}}), in which we have directly
\beq
	\frac{{\rm e}^{-\beta\epsilon_a}}{1+{\rm e}^{-\beta\epsilon_a}} - \frac{{\rm e}^{-\beta\epsilon_b}}{1+{\rm e}^{-\beta\epsilon_b}} =  n(\epsilon_a) - n(\epsilon_b).
\eeq
We obtain the result for the trace of the commutator
\beq
	 \left\langle \left[ S^a_{\bf q}(\tau), S^b_{\bf -q}(0) \right] \right\rangle 
	 = \sum_{{\bf k},s,s'} (\mathcal{S}^a_{\bf k,k+q})_{s,s'} 
(\mathcal{S}^b_{\bf k+q,k})_{s',s} (n(\epsilon_{{\bf k},s}) - n(\epsilon_{{\bf k+q},s'}))   
{\rm e}^{-i\tau(\epsilon_{{\bf k+q},s'}-\epsilon_{{\bf k},s})}.
\eeq
The only time dependent factor is the oscillating exponential and we can thus directly compute the time
 integral in the definition of the susceptibility \eq{\ref{eq chi q_omega def}}
\beqa
	-i\int_{-\infty}^{\infty} {\rm d}\tau {\rm e}^{i\omega \tau} \theta(\tau)
  {\rm e}^{-i\tau(\epsilon_{{\bf k+q},s'}-\epsilon_{{\bf k},s})} &=& 
-i\int_{0}^{\infty} {\rm d}t {\rm e}^{i\omega \tau - \eta \tau} {\rm e}^{-i\tau(\epsilon_{{\bf k+q},s'}-\epsilon_{{\bf k},s})} \nonumber \\
	&=& \frac{1}{\omega + \epsilon_{{\bf k},s'} -\epsilon_{{\bf k+q},s} +i\eta },
\eeqa
where we have added the infinitesimal convergence factor $\eta$. 
Plugging this back to the defining relation for the susceptibility \eq{\ref{eq chi q_omega def}},
we obtain the final result for the susceptibility of the non interacting system
\beq	
		\chi_{\bf q}^{a,b}(\omega) = \frac{1}{N}\sum_{{\bf k},s,s'} (\mathcal{S}^a_{\bf k,k+q})_{s,s'} 
(\mathcal{S}^b_{\bf k+q,k})_{s',s} 
\frac{ n(\epsilon_{{\bf k},s'}) - n(\epsilon_{{\bf k+q},s})}{\omega + \epsilon_{{\bf k},s'} -\epsilon_{{\bf k+q},s} +i\eta}.
	\label{eq chi q_omega}
\eeq

~\\
~\\

\noindent
\emph{Susceptibility in the interacting MF model} \\
We have derived the susceptibility \eq{\ref{eq chi q_omega}} 
for the non-interacting system subjected to a small external driving force. 
We will now use this result to find a susceptibility of the interacting system described by the mean-field theory. 
Lets recall the interaction Hamiltonian \eq{\ref{eq H_int_S}}:
\beqa
	 H_{\rm int} &=& -\frac{2U}{3} \sum_{\bf j} S^b_{\bf j}(t) S^b_{\bf j}(t) =-\frac{2U}{3}\frac{1}{N} \sum_{\bf q} S^b_{\bf -q}(t) S^b_{\bf q}(t) \nonumber \\ 
	 	&\overset{\rm MF}{\equiv}& \frac{-g}{N} \sum_{\bf q}  \langle{S}_{\bf -q}\rangle \langle{S}_{\bf q}\rangle + 
  \langle{S}_{\bf -q}\rangle (S_{\bf q}-\langle{S}_{\bf q}\rangle) + (S_{\bf -q}-\langle{S}_{\bf -q}\rangle) \langle{S}_{\bf q}\rangle + 
O((\delta S)^2) \nonumber \\
	 	&=& \frac{-g}{N} \sum_{\bf q} 2 S_{\bf -q} \langle{S}_{\bf q}\rangle - \langle{S}_{\bf -q}\rangle 
\langle{S}_{\bf q}\rangle = H^{\rm MF}_{\rm int},
\eeqa
where we have introduced the coupling strength $g = 2U/3$. In the last line of the preceding equation, the last term does not contribute to the spin dynamics and is normalized out 
in the computation of the spin average values. We thus drop this term. We obtain the effective external driving Hamiltonian
\beq
	H^{\rm eff}_{\rm ext} = H_{\rm ext} + H^{\rm MF}_{\rm int} = 
\frac{1}{N} \sum_{\bf q} S^b_{\bf -q}(t) \left[ B^b_{\bf q}(t) - 2g\langle{S}^b_{\bf q}\rangle(t) \right] = \frac{1}{N} \sum_{\bf q} S^b_{\bf -q}(t) (B^b_{\bf q})^{\rm eff}(t).
\eeq
We can thus see, that the inclusion of the MF interaction amounts to replacing the external driving $B$ by the new effective driving $B^{\rm eff}$. One then follows the same procedure as in the non interacting case. Since we are mainly interested in the frequency response of the system, we will use the defining formula connecting the frequency components of the spins with the driving through the susceptibility
\beq
	<S^a_{\bf q}(\omega)> = \chi^{a,b}_{\bf q}(\omega) (B^b_{\bf q})^{\rm eff}(\omega).
	\label{eq delta_S avg}
\eeq
One can easily check, that the frequency components of the effective driving are given by the Fourier transform of its parts,
\beqa
	(B^b_{\bf q})^{\rm eff}(\omega) &=& \int {\rm d}t {\rm e}^{i \omega t} (B^b_{\bf q}(t) - 2g \bar{S}^b_{\bf q}(t)) \nonumber \\
		&=& B^b_{\bf q}(\omega) - 2g \bar{S}^b_{\bf q}(\omega).
\eeqa
  Therefore, one obtains
\beq
	\langle{S}^a_{\bf q}\rangle(\omega) = \sum_b \chi^{a,b}_{\bf q}(\omega) B^b_{\bf q}(\omega) - 2g\langle{S}^b_{\bf q}\rangle(\omega),
	\label{eq S avg int}
\eeq 
which after a straightforward manipulation yields the equation for the average value of the spin operators
\beqa
	\langle\mathbf{S}_{\bf q}\rangle(\omega) &=& M^{-1} \chi {\bf B}  \nonumber \\
	M^{a,b} &=& \delta^{a,b} + 2g \chi^{a,b}_{\bf q}(\omega),	
\eeqa
where we merely rewrote the equation \eq{\ref{eq S avg int}} 
in the symbolic matrix notation. From here, one can obtain the information about the critical value of the 
coupling strength $g$ (and thus $U$) from the divergences of the average of the spin operators, 
i.e. when the matrix $M$ becomes singular.




\section{Effective Heisenberg model in the large $U$ limit}
\label{appheis}

We now restrict our interest to the regime with strong repulsion, high $U$. In this regime, the ground state of the grand canonical ensemble has single occupation at each site. Moreover it would cost an energy of order of $U$ to increase the double occupancy by one. This regime can be described by the method of effective Hamiltonian, which is suitable for systems with well separated energy manifolds \cite{Tannoudji_1998}. The energy manifolds are separated by $U$ and we would like to evaluate the effect of the coupling between the ground state manifold and the higher lying manifolds. This coupling results in the perturbation of the bare energy levels in the ground state manifold. In this section, we present the treatment used in \cite{MacDonald_1988} and \cite[page~38]{Tannoudji_1998}.

In the following we consider a situation at half filling $\mu_1 = \mu_2 = 0$. When we multiply the kinetic Hamiltonian \eq{\ref{eq H parts kin}} by $n_{{\bf j},\bar{s}} + h_{{\bf j},\bar{s}} = 1$ from the left and by $n_{{\bf j'},\bar{s'}} + h_{{\bf j'},\bar{s'}} = 1$ from the right, we obtain
\beq
	H_{\rm kin} \equiv T = T_0 + T_{-1} + T_1,
\eeq
where
\beqa
	T_0 &=& - \sum n_{{\bf j},\bar{s}} c^\dag_{{\bf j},s} T_{\bf j,j'}^{s,s'} c_{{\bf j'},s'} n_{{\bf j'}, \bar{s'}}  + h_{{\bf j},\bar{s}} c^\dag_{{\bf j},s} T_{\bf j,j'}^{s,s'} c_{{\bf j'},s'} h_{{\bf j'}, \bar{s'}} \nonumber \\
	T_{-1} &=& -\sum h_{{\bf j},\bar{s}} c^\dag_{{\bf j},s} T_{\bf j,j'}^{s,s'} c_{{\bf j'},s'} n_{{\bf j'}, \bar{s'}}  \nonumber \\
	T_1 &=& -\sum n_{{\bf j},\bar{s}} c^\dag_{{\bf j},s} T_{\bf j,j'}^{s,s'} c_{{\bf j'},s'} h_{{\bf j'}, \bar{s'}},
	\label{eq T}
\eeqa
where $n$ and $h$ denote the particle and hole number operators respectively and $\bar{s}, \bar{s'}$ denote the spin orthogonal to $s, s'$ - e.g. $\bar{s}$ is spin up for $s$ spin down and vice versa. The sums in \eq{\ref{eq T}} run over nearest neighbours $\braket{{\bf j j'}}$ and spins $s,s'$. Denoting the interaction Hamiltonian \eq{\ref{eq H parts int}} as $V$, one can easily verify the following commutation relations
\beqa
	\left[V, T_0 \right] &=& 0 \nonumber \\
	\left[V, T_{-1} \right] &=& -U T_{-1} \nonumber \\
	\left[V, T_1 \right] &=& U T_1,
\eeqa
which can be summarized as
\beq
	\left[V, T_m \right] = m U T_m.
	\label{eq comm VT}
\eeq
We are now ready to proceed with the effective Hamiltonian derivation. We wish to rewrite the current Hamiltonian $H = V + T$ as
\beq
	H_{\rm eff} = {\rm e}^{iS} H {\rm e}^{-iS} = H + \left[iS,H\right] + \frac{1}{2!}\left[iS,\left[ iS, H\right] \right] + \ldots = \sum_{n=0}^\infty \frac{1}{n!} \left[iS,H\right]^{(n)},
	\label{eq H_eff}
\eeq
where $S$ is some Hermitian matrix (we require the transformation to be unitary). $\left[iS,H\right]^{(n)}$ denotes the $n$-times nested commutator and $\left[iS,H\right]^{(0)} = H$. The matrix $S$ can be determined in the following way. Lets write $S$ as
\beq
	S = \lambda S_1 + \lambda^2 S_2 + \ldots = \sum_{n=1}^{\infty} \lambda^n S_n, 
\eeq
where $\lambda = 1/U$ is our small parameter around which we will do our perturbative expansion. The elements of the matrix $S_k$ can be determined in an iterative way by requiring, that after the unitary transformation up to the order $k$ in the expansion \eq{\ref{eq H_eff}} all terms bringing the energy out of the ground state manifold have to vanish. We also denote
\beq
	S^{(k)} = \sum_{n=1}^k \lambda^n S_n
\eeq
Important thing to note is that the energy ratio between $V$ and $T$ is of order $\lambda^{-1}$. It is thus more transparent to rewrite the Hamiltonian as $H = V + \lambda \tilde{T}$, where $V$ and $\tilde{T} = \lambda^{-1} T$ are now contributions of the same order. With this notation, the commutator \eq{\ref{eq comm VT}} can be written as
\beq
	\lambda \left[V, \tilde{T}_m \right] = m \tilde{T}_m.
\eeq
In the first order in $\lambda$ we have from \eq{\ref{eq H_eff}}
\beq
	H_{\rm eff} = H + \left[i\lambda S_1,H\right] = V + \lambda (\tilde{T}_0 + \tilde{T}_{-1} + \tilde{T}_1) + \lambda \left[i S_1, V \right] + \ldots.
\eeq
The terms which bring one from the ground state manifold are the terms changing the double occupancy, i.e. $\tilde{T}_{-1}$ and $\tilde{T}_1$. In order to cancel these terms with the commutator, one gets
\beq
	i S_1 = \lambda (\tilde{T}_1 - \tilde{T}_{-1}). 
\eeq
We will now generalize this procedure in an iterative way to arbitrarily high order in $\lambda$. Lets define a Hamiltonian of order $k+1$ as
\beq
	H^{(k+1)} = {\rm e}^{iS^{(k)}} H {\rm e}^{-iS^{(k)}} = \sum_{n=0}^{\infty} \frac{1}{n!} \left[iS^{(k)},H\right]^{(n)}.
\eeq
As a matter of example, lets take $k=1$
\beqa
	H^{(2)} &=& H + \lambda \left[iS_1, H \right] + \frac{\lambda^2}{2}\lambda \left[iS_1, \left[iS_1, H \right] \right.  + O(\lambda^3)\nonumber \\
	&=& H + \lambda \left[iS_1, V \right] + \lambda^2 \left[iS_1, \tilde{T} \right] + \frac{\lambda^2}{2} \left[iS_1, \left[iS_1, V \right] \right. + O(\lambda^3).
\eeqa
The last two terms in the second line are of the same order $\lambda$ since $S_1 \propto 1$ and $V \propto \lambda^{-1}$. The idea is now to use this Hamiltonian to find the explicit form of the next elements in the expansion of $iS$, namely the element $iS_2$. This can be done as follows. The third order Hamiltonian can be written as
\beq
	H^{(3)} = H^{(2)}_0 + H^{(2)}_{\rm ch} + \lambda^2 \left[iS_2,V\right] + O(\lambda^3),
\eeq
where we have decomposed the second order Hamiltonian into a part which does not change the double occupancy $H^{(2)}_0$ and the one which changes the double occupancy, $H^{(2)}_{\rm ch}$. We require, that the lowest order term of $iS_2$ cancels the double occupancy changing term. This is the general procedure for an arbitrary order expansion, so that
\beq
	H^{(k+1)} = H^{(k)}_0 + H^{(k)}_{\rm ch} + \lambda^k \left[iS_k,V\right] + O(\lambda^{k+1}).
	\label{eq Ham iS_k}
\eeq
In order to find the explicit form of $iS_k$, we will need the following relationships. Lets denote the product of tunnelling operators as
\beq
	\tilde{T}^k(m) \equiv \tilde{T}_{m_1}\ldots \tilde{T}_{m_k}.
\eeq
The commutator with $V$ then reads
\beq
	\lambda \left[ V, \tilde{T}^k(m) \right] = M^k(m) \tilde{T}^k(m),
\eeq
where $M^k(m) = \sum_i m_i$. Looking at \eq{\ref{eq Ham iS_k}}, one can write the last two terms as
\beq
	H^{(k)}_{\rm ch} + \lambda^k \left[iS_k,V\right] = \lambda^{2k-1} \sum_{m} C^k(m) \tilde{T}^k(m) + \lambda^k \left[iS_k,V\right] = 0.	
\eeq
The last equality can be achieved by noting that
\beq
	\lambda \left[ V, \tilde{T}^k(m) \right] \cdot \lambda^{2k-1} \frac{C^k(m)}{M^k(m)} = \lambda^{2k-1} C^k(m) \tilde{T}^k(m),
\eeq
so that we finally obtain
\beq
	iS_k = \lambda^k \sum_m \frac{C^k(m)}{M^k(m)} \tilde{T}^k(m).
\eeq
Note that since $M^k(m) \neq 0$ for double occupancy changing $T^k$, we can safely divide by it.\\
~\\
We have implemented the above described iterative procedure in Mathematica for general tunnellings $T_{\bf j, j'}^{s,s'}$. Up to the second order in the $1/U$ expansion, one obtains the following effective Heisenberg Hamiltonian ($\delta = x,y$ are the spatial directions in the 2D lattice)
\[
 H = H_c + \sum_{\delta=x,y} \sum_{<i,i+\delta>} \sum_{a=x,y,z} J^a_\delta S^a_i S^a_{i+\delta} + 
 \mathbf{D}_{\delta +} \cdot (\mathbf{S}_i \times \mathbf{S}_{i+\delta})_+ + \mathbf{D}_{\delta -} \cdot (\mathbf{S}_i \times \mathbf{S}_{i+\delta})_-
 ,
\]
where
\beqa
	(\mathbf{S}_i \times \mathbf{S}_{i+\delta})_+ &=& (S^y_1 S^z_2, S^z_1 S^x_2, S^x_1 S^y_2) \nonumber \\
	(\mathbf{S}_i \times \mathbf{S}_{i+\delta})_- &=& -(S^z_1 S^y_2, S^x_1 S^z_2, S^y_1 S^x_2) \nonumber
\eeqa
are the "positive" and "negative" part of the vector product. In the most general case, the coefficients read:
\beqa
 J^x_\delta &=& 2 \lambda  \left(T_{\delta}^{2,2} {T_{\delta}^{1,1}}^*+T_{\delta}^{1,2} {T_{\delta}^{2,1}}^*+T_{\delta}^{2,1}
   {T_{\delta}^{1,2}}^*+T_{\delta}^{1,1} {T_{\delta}^{2,2}}^*\right) \nonumber \\   
 J^y_\delta &=& 2 \lambda  \left(T_{\delta}^{2,2} {T_{\delta}^{1,1}}^*-T_{\delta}^{1,2} {T_{\delta}^{2,1}}^*-T_{\delta}^{2,1}
   {T_{\delta}^{1,2}}^*+T_{\delta}^{1,1} {T_{\delta}^{2,2}}^*\right) \nonumber \\   
 J^z_\delta &=& 2 \lambda  \left(T_{\delta}^{1,1} {T_{\delta}^{1,1}}^*-T_{\delta}^{2,1} {T_{\delta}^{2,1}}^*-T_{\delta}^{1,2}
   {T_{\delta}^{1,2}}^*+T_{\delta}^{2,2} {T_{\delta}^{2,2}}^*\right) \nonumber   
\eeqa

\beqa
	D^1_{\delta+} &=& 2 i \lambda  \left(-T_{\delta}^{2,1} {T_{\delta}^{1,1}}^*+T_{\delta}^{1,1} {T_{\delta}^{2,1}}^*+T_{\delta}^{2,2}
   {T_{\delta}^{1,2}}^*-T_{\delta}^{1,2} {T_{\delta}^{2,2}}^*\right) \nonumber \\
	D^2_{\delta+} &=&	2 \lambda  \left(T_{\delta}^{1,2} {T_{\delta}^{1,1}-}^*T_{\delta}^{2,2} {T_{\delta}^{2,1}}^*+T_{\delta}^{1,1}
   {T_{\delta}^{1,2}}^*-T_{\delta}^{2,1} {T_{\delta}^{2,2}}^*\right) \nonumber \\
  D^3_{\delta+} &=&	2 i \lambda  \left(T_{\delta}^{2,2} {T_{\delta}^{1,1}}^*+T_{\delta}^{1,2} {T_{\delta}^{2,1}}^*-T_{\delta}^{2,1}
   {T_{\delta}^{1,2}}^*-T_{\delta}^{1,1} {T_{\delta}^{2,2}}^*\right) \nonumber
\eeqa

\beqa	
	D^1_{\delta-} &=& 2 i \lambda  \left(-T_{\delta}^{1,2} {T_{\delta}^{1,1}}^*+T_{\delta}^{2,2} {T_{\delta}^{2,1}}^*+T_{\delta}^{1,1}
   {T_{\delta}^{1,2}}^*-T_{\delta}^{2,1} {T_{\delta}^{2,2}}^*\right) \nonumber \\
  D^2_{\delta-} &=& 2 \lambda  \left(-T_{\delta}^{2,1} {T_{\delta}^{1,1}}^*-T_{\delta}^{1,1} {T_{\delta}^{2,1}}^*+T_{\delta}^{2,2}
   {T_{\delta}^{1,2}}^*+T_{\delta}^{1,2} {T_{\delta}^{2,2}}^*\right) \nonumber \\
  D^3_{\delta-} &=& 2 i \lambda  \left(T_{\delta}^{2,2} {T_{\delta}^{1,1}}^*-T_{\delta}^{1,2} {T_{\delta}^{2,1}}^*+T_{\delta}^{2,1}
   {T_{\delta}^{1,2}}^*-T_{\delta}^{1,1} {T_{\delta}^{2,2}}^*\right) \nonumber 
\eeqa

\[
	H_c = 2 \lambda  \left(-T_{\delta}^{1,1} {T_{\delta}^{1,1}}^*-T_{\delta}^{2,1} {T_{\delta}^{2,1}}^*-T_{\delta}^{1,2}
   {T_{\delta}^{1,2}}^*-T_{\delta}^{2,2} {T_{\delta}^{2,2}}^*\right) \frac{\mathds{1}}{4}
\]


\section{Magnetic orders}
\label{sec App mag orders}

In Table \ref{tab order pars} we list the details of the magnetic orders parametrized by \eq{\ref{eq mag order}} and summarized in Table \ref{tab phase orderings}. The tables have the following format


\[\begin{array}{l}

\begin{array}{c}
 \text{$\{\alpha, U, \beta \}$}, {\rm Phase}
\end{array}
 \\

\begin{array}{|c|c|c|c|c|c|c|c|}
\hline
 q_x/\pi & q_y/\pi & N_1^x & N_2^x & N_1^y & N_2^y & N_1^z & N_2^z \\
 \hline 
\end{array}
 \\
 
\end{array}\]

\newpage

\begin{table}
\begin{tabular}{c}
$
\begin{array}{l}
 
\begin{array}{c}
 \text{$\{\alpha, U, \beta \}$} = \text{$\{$0.1, 2.0, 10000$\}$}, {\rm AF}
\end{array}
 \\
 
\begin{array}{|c|c|c|c|c|c|c|c|}
\hline
 q_x/\pi & q_y/\pi & N_1^x & N_2^x & N_1^y & N_2^y & N_1^z & N_2^z \\ \hline
 -1 & 1 & 1.8\times 10^{-2} & 0 & -1.8\times 10^{-2} & 0 & 0 & 0 \\ \hline
 1 & 1 & 1.8\times 10^{-2} & 0 & -1.8\times 10^{-2} & 0 & 0 & 0 \\ \hline
\end{array}
 \\
 
\begin{array}{c}
 \text{}
\end{array}
 \\
 
\begin{array}{c}
 \text{$\{\alpha, U, \beta \}$} = \text{$\{$0.1, 2.5, 10000$\}$}, {\rm SI2}
\end{array}
 \\
 
\begin{array}{|c|c|c|c|c|c|c|c|}
\hline
 q_x/\pi & q_y/\pi & N_1^x & N_2^x & N_1^y & N_2^y & N_1^z & N_2^z \\ \hline
 -1 & \frac{13}{18} & 0 & 0 & -5.3\times 10^{-2} & 8.5\times 10^{-2} & -9.4\times 10^{-2} & -5.8\times 10^{-2} \\ \hline
 1 & \frac{13}{18} & 0 & 0 & -5.3\times 10^{-2} & 8.5\times 10^{-2} & -9.4\times 10^{-2} & -5.8\times 10^{-2} \\ \hline
\end{array}
 \\
 
\begin{array}{c}
 \text{}
\end{array}
 \\
 
\begin{array}{c}
 \text{$\{\alpha, U, \beta \}$} = \text{$\{$0.2, 3.0, 1000$\}$}, {\rm AF}
\end{array}
 \\
 
\begin{array}{|c|c|c|c|c|c|c|c|}
\hline
 q_x/\pi & q_y/\pi & N_1^x & N_2^x & N_1^y & N_2^y & N_1^z & N_2^z \\ \hline
 -1 & 1 & 2.1\times 10^{-2} & 0 & -2.1\times 10^{-2} & 0 & 0 & 0 \\ \hline
 1 & 1 & 2.1\times 10^{-2} & 0 & -2.1\times 10^{-2} & 0 & 0 & 0 \\ \hline
\end{array}
 \\
 
\begin{array}{c}
 \text{}
\end{array}
 \\
 
\begin{array}{c}
 \text{$\{\alpha, U, \beta \}$} = \text{$\{$0.2, 3.5, 50$\}$}, {\rm SI}
\end{array}
 \\
 
\begin{array}{|c|c|c|c|c|c|c|c|}
\hline
 q_x/\pi & q_y/\pi & N_1^x & N_2^x & N_1^y & N_2^y & N_1^z & N_2^z \\ \hline
 -\frac{5}{6} & \frac{5}{9} & -4.4\times 10^{-3} & -2.1\times 10^{-2} & 9.3\times 10^{-3} & 4.5\times 10^{-2} & -4.2\times 10^{-2} & 8.8\times 10^{-3} \\ \hline
 \frac{5}{6} & \frac{5}{9} & -2.2\times 10^{-2} & 4.5\times 10^{-4} & -4.6\times 10^{-2} & 9.4\times 10^{-4} & -8.9\times 10^{-4} & -4.3\times 10^{-2} \\ \hline
\end{array}
 \\
 
\begin{array}{c}
 \text{}
\end{array}
 \\
 
\begin{array}{c}
 \text{$\{\alpha, U, \beta \}$} = \text{$\{$0.2, 4.0, 10000$\}$}, {\rm SI}
\end{array}
 \\
 
\begin{array}{|c|c|c|c|c|c|c|c|}
\hline
 q_x/\pi & q_y/\pi & N_1^x & N_2^x & N_1^y & N_2^y & N_1^z & N_2^z \\ \hline
 -\frac{1}{2} & \frac{8}{9} & 1.1\times 10^{-1} & 2.9\times 10^{-2} & -3.8\times 10^{-2} & -1.\times 10^{-2} & 2.7\times 10^{-2} & -1.\times 10^{-1} \\ \hline
 \frac{1}{2} & \frac{8}{9} & 2.9\times 10^{-2} & -1.1\times 10^{-1} & 1.\times 10^{-2} & -3.8\times 10^{-2} & 1.\times 10^{-1} & 2.7\times 10^{-2} \\ \hline
\end{array}
 \\
 
\begin{array}{c}
 \text{}
\end{array}
 \\
 
\begin{array}{c}
 \text{$\{\alpha, U, \beta \}$} = \text{$\{$0.2, 4.0, 50$\}$}, {\rm SI}
\end{array}
 \\
 
\begin{array}{|c|c|c|c|c|c|c|c|}
\hline
 q_x/\pi & q_y/\pi & N_1^x & N_2^x & N_1^y & N_2^y & N_1^z & N_2^z \\ \hline
 -\frac{1}{2} & \frac{8}{9} & 1.1\times 10^{-1} & 2.9\times 10^{-2} & -3.8\times 10^{-2} & -1.\times 10^{-2} & 2.7\times 10^{-2} & -1.\times 10^{-1} \\ \hline
 \frac{1}{2} & \frac{8}{9} & 2.9\times 10^{-2} & -1.1\times 10^{-1} & 1.\times 10^{-2} & -3.8\times 10^{-2} & 1.\times 10^{-1} & 2.7\times 10^{-2} \\ \hline
\end{array}
 \\
 
\begin{array}{c}
 \text{}
\end{array}
 \\
 
 
\begin{array}{c}
 \text{$\{\alpha, U, \beta \}$} = \text{$\{$0.25, 3.5, 1000$\}$}, {\rm AF}
\end{array}
 \\
 
\begin{array}{|c|c|c|c|c|c|c|c|}
\hline
 q_x/\pi & q_y/\pi & N_1^x & N_2^x & N_1^y & N_2^y & N_1^z & N_2^z \\ \hline
 -1 & 1 & 1.6\times 10^{-2} & 0 & -4.4\times 10^{-2} & 0 & 0 & 0 \\ \hline
 1 & 1 & 1.6\times 10^{-2} & 0 & -4.4\times 10^{-2} & 0 & 0 & 0 \\ \hline
\end{array}
 \\
 
\begin{array}{c}
 \text{}
\end{array}
 \\
 
\begin{array}{c}
 \text{$\{\alpha, U, \beta \}$} = \text{$\{$0.25, 4.0, 50$\}$}, {\rm SI}
\end{array}
 \\
 
\begin{array}{|c|c|c|c|c|c|c|c|}
\hline
 q_x/\pi & q_y/\pi & N_1^x & N_2^x & N_1^y & N_2^y & N_1^z & N_2^z \\ \hline
 -\frac{7}{18} & \frac{7}{9} & -4.\times 10^{-2} & 1.2\times 10^{-2} & 1.8\times 10^{-2} & -5.5\times 10^{-3} & 1.1\times 10^{-2} & 3.8\times 10^{-2} \\ \hline
 \frac{7}{18} & \frac{7}{9} & -2.8\times 10^{-2} & 3.1\times 10^{-2} & -1.3\times 10^{-2} & 1.4\times 10^{-2} & -2.9\times 10^{-2} & -2.6\times 10^{-2} \\ \hline
\end{array}
 \\
 
\begin{array}{c}
 \text{}
\end{array}
 \\
 \end{array}
$
\end{tabular}
\end{table}

\begin{table}
\begin{tabular}{c}
$
\begin{array}{l}
\begin{array}{c}
 \text{$\{\alpha, U, \beta \}$} = \text{$\{$0.3, 4.5, 50$\}$}, {\rm AF}
\end{array}
 \\
 
\begin{array}{|c|c|c|c|c|c|c|c|}
\hline
 q_x/\pi & q_y/\pi & N_1^x & N_2^x & N_1^y & N_2^y & N_1^z & N_2^z \\ \hline
 -1 & 1 & 0 & 0 & -6.1\times 10^{-2} & 0 & 0 & 0 \\ \hline
 1 & 1 & 0 & 0 & -6.1\times 10^{-2} & 0 & 0 & 0 \\ \hline
\end{array}
 \\
 
\begin{array}{c}
 \text{}
\end{array}
 \\
 
\begin{array}{c}
 \text{$\{\alpha, U, \beta \}$} = \text{$\{$0.3, 4.75, 50$\}$}, {\rm SI}
\end{array}
 \\
 
\begin{array}{|c|c|c|c|c|c|c|c|}
\hline
 q_x/\pi & q_y/\pi & N_1^x & N_2^x & N_1^y & N_2^y & N_1^z & N_2^z \\ \hline
 \frac{2}{9} & \frac{13}{18} & -5.9\times 10^{-2} & -3.2\times 10^{-2} & -1.9\times 10^{-2} & -1.\times 10^{-2} & -2.5\times 10^{-2} & 4.6\times 10^{-2} \\ \hline
 \frac{13}{18} & \frac{2}{9} & -2.1\times 10^{-2} & 6.2\times 10^{-3} & -6.4\times 10^{-2} & 1.9\times 10^{-2} & 1.5\times 10^{-2} & 5.\times 10^{-2} \\ \hline
\end{array}
 \\
 
\begin{array}{c}
 \text{}
\end{array}
 \\
 
\begin{array}{c}
 \text{$\{\alpha, U, \beta \}$} = \text{$\{$0.3, 12.0, 50$\}$}, {\rm SkX}
\end{array}
 \\
 
\begin{array}{|c|c|c|c|c|c|c|c|}
\hline
 q_x/\pi & q_y/\pi & N_1^x & N_2^x & N_1^y & N_2^y & N_1^z & N_2^z \\ \hline
 -1 & \frac{1}{3} & 0 & 0 & -2.8\times 10^{-1} & 1.6\times 10^{-1} & 6.9\times 10^{-2} & 1.2\times 10^{-1} \\ \hline
 -\frac{1}{3} & 1 & -2.8\times 10^{-1} & -1.6\times 10^{-1} & 0 & 0 & 6.9\times 10^{-2} & -1.2\times 10^{-1} \\ \hline
 \frac{1}{3} & 1 & -2.8\times 10^{-1} & 1.6\times 10^{-1} & 0 & 0 & 6.9\times 10^{-2} & 1.2\times 10^{-1} \\ \hline
 1 & \frac{1}{3} & 0 & 0 & -2.8\times 10^{-1} & 1.6\times 10^{-1} & 6.9\times 10^{-2} & 1.2\times 10^{-1} \\ \hline
\end{array}

\end{array}
$
\end{tabular}

\caption{Magnetic order parameters ${\bf N_i} = (N_i^x,N_i^y,N_i^z)$, $i=1,2$ 
introduced in \eq{\ref{eq mag order}} and the peak values of ${\bf q} = (q_x,q_y)$, together with the parameters $\{\alpha, U, \beta \}$  for phases given in Table \ref{tab Summary}.
In the table, we provide the data for magnetic order parameters only for $q_y>0$, since the values for $q_y<0$ are related by
the inversion symmetry ${\bf{q}}\rightarrow{-\bf{q}}$.}
\label{tab order pars}

\end{table}

\end{widetext}

\end{document}